\documentstyle[multicol,aps,epsf]{revtex} 
\begin{document}

\title{Inhomogeneous Steady States of Diffusion-Limited Coalescence,
$A+A\rightleftharpoons A$}
\author{Daniel ben-Avraham\footnote
  {{\bf e-mail:} qd00@polaris.clarkson.edu}}

\address{Physics Department, and Clarkson Institute for Statistical
Physics (CISP), \\ Clarkson University, Potsdam, NY 13699-5820}
\maketitle

\begin{abstract} 
We study the steady state of diffusion-limited coalescence,
$A+A\rightleftharpoons A$, in the presence of a trap and with a background
drift.  In one dimension this model can be analyzed exactly through the method
of inter-particle distribution functions (IPDF).  Because of the
irreversible trap the steady state of the system is a non-equilibrium 
state.  An interesting phase
transition, controlled by the drift away from the trap, takes place: from a
non-trivial steady state, when the drift is weak, to a trivial steady state
(the vacuum), as the drift increases beyond some critical point.  Surprisingly,
regardless of the drift strength, the computed IPDF resembles that of an
homogeneous {\em equilibrium\/} system, without the trap. We suggest that this
is due to ``shielding": the particle nearest to the trap shields the remaining
particles from the effects of the trap. Finally, we compare the exact solution
to that of a reaction-diffusion equation, and we determine the optimal
values of the appropriate rate coefficients.
\end{abstract} 

\begin{multicols}{2}

\section{Introduction }
\label{sec:intro} 
Non-equilibrium kinetics of diffusion-limited reactions has been the subject
of much recent interest
\cite{vanKampen,Haken,Nicolis,Ligget,Rice,reaction-reviews,JStatPhys}.  In
contrast to equilibrium systems---which are best analyzed with standard
thermodynamics---or reaction-limited processes---whose kinetics is well
described by classical rate equations \cite{Laidler,Benson}---there is no
general approach to non-equilibrium, diffusion-limited reactions. 

In this paper we study a diffusion-limited coalescence process in one
dimension: $A+A\rightleftharpoons A$, which yields itself to {\em exact\/}
analysis
\cite{Coalescence,DbArevs,Doering,Krebs,Henkel,Simon,Droz,Bramson,%
Torney,Spouge,Takayasu,Privman}.  We show that a solution is possible even
in the presence of background drift, and in the presence of a trap; when in the
long-time asymptotic limit the stationary state of the system is a {\em
nonequilibrium\/} state.  The drift field controls an interesting phase
transition in the steady state of the system:  The size of the depletion zone
next to the trap increases as a function of the drift away from the trap. 
Beyond a certain critical drift strength, the only possible stationary state is
the vacuum (an empty lattice, with no particles).

Our analysis of the steady state yields the surprising result that the
distance between a given particle and the nearest particle to its right (the
``forward" IPDF) is exactly the same as in equilibrium, even in the presence of
a trap, and independently of the drift!  We suggest that this
is due to a ``shielding" effect: the particle nearest to the trap shields the
rest of the particles from the effect of the trap.  The trap
and the drift influence only the distance between the trap
and the nearest particle to the trap---all other particles remain unaffected.

Finally, we use the coalescence model with a trap to study the applicability of
reaction-diffusion equations.  Reaction-diffusion equations are the most common
approximation method in the study of diffusion-limited processes.  Our system
is a rare example where (in the absence of drift) the stationary solution of
the relevant reaction-diffusion equation may be found in closed form.  We
explore in what ways the reaction-diffusion  equation is approximate, by
comparing its results  to the exact solution of the IPDF method.  The
comparison also allows us to determine the optimal rate constants of the
reaction-diffusion equation in a straightforward manner.  Normally, this feat
requires a renormalization group analysis.

The rest of this paper is organized as follows. In
Section~\ref{sec:formulation} we present a lattice model of diffusion-limited
reversible coalescence, along with the exact method of analysis; the method of
Empty Intervals, also known as the method of Inter-Particle Distribution
Functions (IPDF). The stationary {\em equilibrium\/} state in a homogeneous
infinite system, needed for comparison with the trap, is analyzed and
summarized in Section~\ref{sec:homogeneous}.  In Section~\ref{sec:trap}, a trap
is introduced and the resulting stationary state is explored.  The exact
solution includes a description of the dynamical phase transition in the
stationary kinetics of the system, brought about by the background drift.  The
explanation to the surprising result that the IPDF of the system with a trap is
homogeneous (and exactly the same as in equilibrium) is presented in
Section~\ref{sec:shielding}.  Last, in Section~\ref{sec:mean-field} we
compare the exact solution for the case of non-biased diffusion (zero drift)
to that of the corresponding reaction-diffusion equation.  Our goal is to show
in what ways the latter is an approximation, and to devise strategies to find
out appropriate rate coefficients.  We conclude with a discussion and open
questions in Section~\ref{sec:discussion}.

\section{Reversible Coalescence} 
\label{sec:formulation}

Our model \cite{Coalescence,DbArevs,Doering} is defined on a one-dimensional
lattice of lattice spacing
$a$.  Each site is in one of two states: occupied by a particle $A$
($\bullet$), or empty ($\circ$). Particles hop randomly to the nearest neighbor
site on their right (left), at rate $D/a^2+u/2a$\ $(D/a^2-u/2a)$.  Thus, in the
continuum limit of
$a\to 0$, the particles undergo diffusion with a diffusion constant $D$, and
with a uniform background drift velocity $u$.  A particle may give birth to an
additional particle, into a nearest neighbor site, at rate
$v/a$ (on either side of the particle) \cite{remark}.  If hopping or birth
occurs into a site which is already occupied, the target site remains
occupied.  The last rule means that coalescence,
$A+A\to A$, takes place {\em immediately\/} upon encounter of any two
particles.  Thus, the system models the
diffusion-limited reaction process
\begin{equation}
A+A\rightleftharpoons A\;. 
\end{equation}
The dynamical rules of the model are illustrated in Fig.~1.

\begin{figure}
\centerline{\epsfxsize=5cm \epsfbox{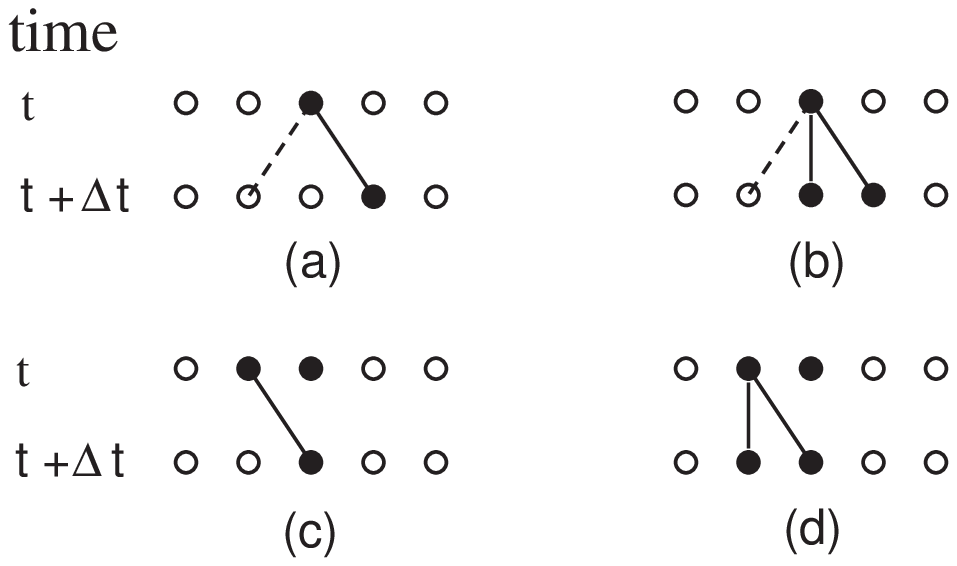}}
\noindent
{\small {\bf Figure~1}. Reaction rules: (a) diffusion; (b)
birth; and coalescence, (c) following diffusion, and (d) following a birth
event.  The dotted lines in (a) and (b) indicate alternative target sites.  In
(a), the rate of the two possibilities differ due to the drift.}
\label{fig:rules}
\end{figure} 
 
An exact treatment of the problem is possible through the method of Empty
Intervals, known also as the method of Inter-Particle Distribution Functions
(IPDF).  The key concept is
$E_{n,m}(t)$---the probability that sites $n,n+1,\cdots,m$ are empty at time
$t$.  The probability that site $n$ is occupied is 
\begin{equation}
\label{disc-conc}
{\rm Prob}({\rm site\ }n{\rm\ is\ occupied})
\equiv{\rm Prob}\big(\stackrel{n}{\bullet}\big)=1-E_{n,n}\;.
\end{equation}
The event that sites $n$ through $m$ are empty (prob. $E_{n,m}$) consists of two
cases: site $m+1$ is also empty (prob. $E_{n,m+1}$), or it is occupied.  Thus
the probability that sites $n$ through $m$ are empty, but site
$m+1$ is occupied is 
\begin{equation}
\label{edge1}
{\rm Prob}(\stackrel{n}{\circ}\cdots\stackrel{m}{\circ}\bullet)
=E_{n,m}-E_{n,m+1}\;,  
\end{equation}
and, likewise,
\begin{equation}
\label{edge2}
{\rm Prob}(\bullet\stackrel{n}{\circ}\cdots\stackrel{m}{\circ})
=E_{n,m}-E_{n-1,m}\;.  
\end{equation}
With this in mind, one can write down
a rate equation for the evolution of the empty interval probabilities:
\begin{eqnarray}
\label{mastereq}
&&{\partial E_{n,m}\over\partial t}
   = ({D\over a^2}+{u\over2a})(E_{n,m-1}-E_{n,m})\nonumber\\
   &&- ({D\over a^2}-{u\over2a})(E_{n,m}-E_{n,m+1})\nonumber\\
   &&- ({D\over a^2}+{u\over2a})(E_{n,m}-E_{n-1,m})\nonumber\\
   &&+ ({D\over a^2}-{u\over2a})(E_{n+1,m}-E_{n,m})\nonumber\\
   &&- {v\over a}[(E_{n,m}-E_{n,m+1})+(E_{n,m}-E_{n-1,m})]\;.
\end{eqnarray}
For example, the first term on the r.h.s. of Eq.~(\ref{mastereq}) accounts for
the increase in $E_{n,m}$ when the particle at the right edge of
$\,\stackrel{n}{\circ}\cdots\circ\!\stackrel{m}\bullet\,$ hops to the right 
and the sites $n,\dots,m$ become empty; the second term denotes the decrease
in $E_{n,m}$ when a particle at $m+1$ hops to the left into the empty interval
$n,\dots,m$, and so on. 

Eq.~(\ref{mastereq}) is valid for
$m>n$.  The special case of
$m=n$ corresponds to $E_{n,n}$---the probability that site $n$ is empty.  It
is described by the equation
\begin{eqnarray}
\label{mastereq_n,n}
&&{\partial E_{n,n}\over\partial t}
   = ({D\over a^2}+{u\over2a})(1-E_{n,n})\nonumber\\
   &&- ({D\over a^2}-{u\over2a})(E_{n,n}-E_{n,n+1})\nonumber\\
   &&- ({D\over a^2}+{u\over2a})(E_{n,n}-E_{n-1,n})\nonumber\\
   &&+ ({D\over a^2}-{u\over2a})(1-E_{n,n})\nonumber\\
   &&- {v\over a}[(E_{n,n}-E_{n,n+1})+(E_{n,n}-E_{n-1,n})]\;.
\end{eqnarray}
Comparison with Eq.~(\ref{mastereq}) yields the boundary condition:
\begin{equation}
\label{discBC}
E_{n,n-1}=1\;.
\end{equation}
The fact that the $\{E_{n,m}\}$ represent {\em probabilities\/} implies the
additional condition that 
\begin{equation}
E_{n,m}\geq 0\;.
\end{equation}  
Finally, if the system is not empty
then 
\begin{equation}
\lim_{{n\to-\infty\atop m\to+\infty}}E_{n,m}= 0\;.  
\end{equation}

In many applications, it is simpler to pass to the continuum limit.  We write
$x=na$ and $y=ma$, and replace $E_{n,m}(t)$ with $E(x,y;t)$.  Letting $a\to 0$,
Eq.~(\ref{mastereq}) becomes
\begin{equation}
\label{eqE}
{\partial E\over\partial t}=D({\partial^2\over\partial x^2}
  +{\partial^2\over\partial y^2})E - (u+v){\partial E\over\partial x}
  -(u-v){\partial E\over\partial y}\;,
\end{equation}
with the boundary conditions, 
\begin{eqnarray}
\label{BC1}
E(x,x;t) &=& 1\;,\\
\label{BCpositive}
E(x,y;t) &\geq& 0\;,\\
\label{BC0}
\lim_{{x\to-\infty\atop y\to+\infty}}E(x,y;t) &=& 0\;.
\end{eqnarray}

The concentration of particles is obtained from the continuum limit of
Eq.~(\ref{disc-conc}), together with Eq.~(\ref{BC1}):
\begin{equation}
\label{conc}
c(x,t)=-{\partial E(x,y;t)\over\partial y}|_{y=x}\;.
\end{equation} 
The conditional joint probability for having particles at
$n$ and $m$ but none in between, is (Fig.~2)
\begin{eqnarray}
\label{disc-IPDF}
P_{n,m}(t)&=&
   {\rm Prob}(\stackrel{n}{\bullet}\circ\cdots\circ\stackrel{m}{\bullet})
      \nonumber\\
&=&E_{n+1,m-1}-E_{n+1,m}-E_{n,m-1}+E_{n,m}\;,
\end{eqnarray}
which in the continuum limit becomes
\begin{equation}
\label{IPDF}
P(x,y;t)=-{\partial^2E(x,y;t)\over\partial x\,\partial y}\;.
\end{equation}
{}From $P$ we can compute the IPDF\@.  The ``forward" IPDF, {\it i.e.}, the
probability that given a particle at
$x$ the next nearest particle {\em to its right\/} is at $y$, is 
\begin{equation}
\label{IPDFf}
p(x,y;t)=c(x,t)^{-1}P(x,y;t)\;.
\end{equation}
Likewise, the ``backward" IPDF---the probability that given a
particle at $y$ the next nearest particle {\em to its left\/} is at $x$---is
\begin{equation}
\label{IPDFb}
q(x,y;t)=c(y,t)^{-1}P(x,y;t)\;.
\end{equation}

\begin{figure}
\centerline{\epsfxsize=5cm \epsfbox{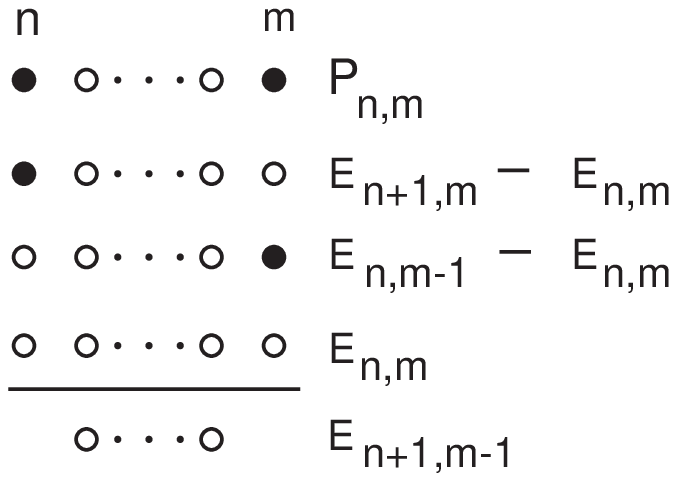}}
\noindent
{\small{\bf Figure~2}. Discrete IPDF: $E_{n+1,m-1}$ consists of four
events, listed above the ``total" line.  In this way one can find
$P_{n,m}$---the only remaining unknown.}
\label{fig:IPDF}
\end{figure}
 
\section{The equilibrium state}
\label{sec:homogeneous}
We now derive the steady state solution of Eq.~(\ref{eqE}), with 
the boundary conditions~(\ref{BC1}) -- (\ref{BC0}).  It proves useful to
change variables: $\eta=x+y$, and $\xi=y-x$.  $\xi$ represents the length
of the intervals, and $\eta$ represents their location (the center of the
interval is at
$\eta/2$).  In these new variables, the stationary limit of Eq.~(\ref{eqE})
becomes
\begin{equation}
\label{eqEs}
0=({\partial^2\over\partial\eta^2}+{\partial^2\over\partial\xi^2})E
-\omega{\partial\over\partial\eta}E+\gamma{\partial\over\partial\xi}E\;,
\end{equation}
where we have used the shorthand notation $\omega\equiv u/D$ and $\gamma\equiv
v/D$.  

Because the background drift is homogeneous, we expect the solution to
be translationally invariant, that is, the probability of empty
intervals should depend only on their length but not upon their
location: $E=E(\xi)$. $E$ then satisfies a simple ODE;
$d^2E/d\xi^2-\gamma dE/d\xi=0$, whose solution (in view of the boundary
conditions) is 
\begin{equation}
\label{E_eq}
E_{eq}=e^{-\gamma(y-x)}\;.
\end{equation}
This corresponds to the particle concentration (Eq.~\ref{conc})
\begin{equation}
\label{ceq}
c_{eq}=\gamma={v\over D}\;.
\end{equation}
The subscript ``eq" emphasizes the fact that here the stationary
solution is an {\em equilibrium\/} solution.  This is to be expected because
all the microscopic rules involved are reversible.  Notice that the
equilibrium solution is independent of the drift.  The effect
of the drift is to impart the whole system a uniform background velocity,
$u$,  which in fact may be eliminated through a Galilean transformation. 

We can test that the distribution of particles is indeed an equilibrium
distribution, by working backwards.  At equilibrium the particles ought to be
distributed randomly, independently from each other---a state of maximum
entropy.  Suppose that the particle concentration is $\gamma$, then the
probability that there are no particles in an infinitesimal interval of length
$\Delta\xi$ is
$(1-\gamma\Delta\xi)$.  Since the particles are uncorrelated, the probability
that a finite interval of length $\xi=y-x$ is empty is then
$(1-\gamma\Delta\xi)^{\xi/\Delta\xi}$.  In the limit $\Delta\xi\to 0$, one
recovers the equilibrium empty interval probability, Eq.~(\ref{E_eq}).  

Finally, we can work out the IPDF\@.  Using Eqs.~(\ref{IPDF}) -- (\ref{IPDFb}),
(\ref{E_eq}), and (\ref{ceq}), we find the Poisson distribution: 
\begin{equation}
\label{IPDF.eq}
p_{eq}(x,y)=q_{eq}(x,y)=\gamma e^{-\gamma(y-x)}\;.
\end{equation}
Again, this can be obtained, independently, from an analysis of
uncorrelated, randomly distributed particles.

\section{Steady state with a trap} 
\label{sec:trap}

The equilibrium solution, Eq.~(\ref{E_eq}), is not the only solution to
Eq.~(\ref{eqEs}).  Another trivial solution is the ``vacuum" state; $E(x,y)=1$.
It represents a completely empty system.  But the solution~(\ref{E_eq}) is
stable, while the vacuum state is not.  In fact, when the initial state of the
system is a mixture of the two phases:
$c(x,t\!=\!0)=0$ for $x<0$, and $c(x,t\!=\!0)=c_{eq}$ for $x>0$, say, then the
stable phase invades the unstable phase.  In the absence of drift, the front
between the two phases propagates at the constant speed $v$ (the birth rate)
\cite{Doering2}.
This system has been studied as a ``noisy" analogue of the mean-field problem
of Fisher waves \cite{waves}. Here we wish to study another inhomogeneous
situation, where there is a perfectly absorbing trap at the origin, instead of
the initial empty half-space.  

As opposed to the Fisher problem, in the case
of a trap the system arrives at a stationary state.  This state must be a
non-equilibrium state, because the trap is irreversible (particles trapped in
can't ever come out).  As the trap depletes its immediate neighborhood, the
depletion zone is continually replenished by a stream of
particles from the stable phase which rush in at speed $v$, as in the Fisher
wave problem.  Suppose now that the particles are subject to a drift $u$ away
from the trap, then an interesting transition would take place as $u$ is made
larger than $v$: the depletion zone would grow faster than it could be
replenished, and the stationary state would never be achieved!  We wish to study
this transition, and how the depletion zone is affected by the drift when
$u<v$.

To derive the appropriate boundary condition, we turn back to
the discrete representation.  In the presence of a perfect trap at $n=0$,
Eq.~(\ref{mastereq}) is limited to $0<n<m$.  The special equation for
$n=0$ is
\begin{eqnarray}
{\partial E_{0,m}\over\partial t}
   &=& ({D\over a^2}+{u\over2a})(E_{0,m-1}-E_{0,m})\nonumber\\
   &-& ({D\over a^2}-{u\over2a})(E_{0,m}-E_{0,m+1})\nonumber\\
   &+& ({D\over a^2}-{u\over2a})(E_{1,m}-E_{0,m})\nonumber\\
   &-& {v\over a}(E_{0,m}-E_{0,m+1})\;.
\end{eqnarray} 
Comparison to Eq.~(\ref{mastereq}) yields the discrete boundary condition
\begin{equation}
E_{-1,m}=E_{0,m}\;,
\end{equation}
which in the continuum limit becomes
\begin{equation}
\label{BCtrap}
{\partial E(x,y;t)\over\partial x}|_{x=0}=0\;.
\end{equation}
In addition, the boundary condition~(\ref{BC0}) is now replaced by
\begin{equation}
\label{BC0trap}
\lim_{y\to\infty}E(0,y;t) = 0\;.
\end{equation}
 
We are looking then for the solution to Eq.~(\ref{eqEs}), confined to the wedge
($0\leq x\leq y$), and which satisfies the boundary conditions (\ref{BC1}),
(\ref{BCpositive}), (\ref{BCtrap}), and (\ref{BC0trap}).  Following the method
of separation of variables, assume
$E(\eta,\xi)=\sum_{\alpha}A_{\alpha}f_{\alpha}(\eta)g_{\alpha}(\xi)$, then
$f_{\alpha}$ and $g_{\alpha}$ satisfy the ODEs:
\begin{eqnarray}
&&{d^2f_{\alpha}\over d\eta^2}-\omega{df_{\alpha}\over d\eta}=
  +\alpha^2f_{\alpha}\;,\\
&&{d^2g_{\alpha}\over d\xi^2}+\gamma{dg_{\alpha}\over d\xi}=
  -\alpha^2g_{\alpha}\;.
\end{eqnarray}
We find the solution by inspection: A good guess is that $\alpha^2=0$ would be
part of it, since far away from the trap one expects convergence to
the equilibrium state.  However, $f_0g_0$ fails to satisfy the boundary
condition due to the trap, Eq.~(\ref{BCtrap}).  Looking for the simplest
possible way to mend this and to satisfy the remaining boundary
conditions, we find that it is sufficient to consider just one additional
component: $4\alpha^2=\gamma^2-\omega^2$.  The final solution turns out to be
\begin{equation} 
\label{Es.drift}
E_s(x,y)=e^{-\gamma(y-x)}+{\gamma\over\omega}e^{-\gamma y}(e^{\omega y}
  -e^{\omega x})\;.
\end{equation}
Far away from the trap, as $x,y\to\infty$, this converges to the equilibrium
result of Eq.~(\ref{E_eq}). 

From~(\ref{conc}), we obtain the stationary concentration profile:
\begin{equation}
\label{cs}
c_s(x)=\gamma[1-e^{-(\gamma-\omega)x}]\;.
\end{equation}
There is a depletion zone of size $(\gamma-\omega)^{-1}=D/(v-u)$ near the trap,
and the concentration grows asymptotically to $c_{eq}=\gamma$ as $x\to\infty$. 
As the drift velocity $u$ grows and approaches
$v$, the depletion zone becomes larger and larger.  In the limit $u\to v$ the
depletion zone is infinite: from~(\ref{Es.drift}) we see that then $E_s\to 1$,
{\it i.e.}, the stationary state is the vacuum (but it takes an infinite
time to get there).

The IPDF is surprising.  From~(\ref{IPDF}) we get the conditional
joint probability
\begin{equation}
P_s(x,y)=\gamma^2e^{-\gamma(y-x)}[1-e^{-(\gamma-\omega)x}]\;,
\end{equation}
and so the forward IPDF is
\begin{equation}
\label{ps}
p_s(x,y)=\gamma e^{-\gamma\xi}\;;\qquad \xi\equiv y-x\;,
\end{equation}
regardless of the drift.  The notation chosen here emphasizes the fact
that $p_s$ is translationally invariant, in spite of the trap at the origin.
In fact, the forward IPDF is exactly the same as in the {\em equilibrium\/}
state (Eq.~\ref{IPDF.eq})! 

The backward IPDF is
\begin{equation}
q_s(x,y)=\gamma{e^{-\gamma\xi}-e^{-\omega\xi}e^{-(\gamma-\omega)y}\over
  1-e^{-(\gamma-\omega)y} }\;;\qquad \xi\equiv y-x\;. 
\end{equation} 
Notice this time the dependence on the position of the given particle, $y$, as
well as the dependence on the drift, and that the backward and forward IPDFs
are {\em not\/} equal.  
Moreover, the backward IPDF does not normalize properly. The reason for that is
that  there is a finite
chance that there are no particles between the particle at $y$ and
the trap; {\it i.e.}, the particle at $y$ might be the nearest
particle to the trap.  The probability density for the distance of the nearest
particle to the trap can be computed from the continuum limit of
Eq.~(\ref{edge1});
$p_0(y)=-\partial E/\partial y|_{x=0}$, or 
\begin{equation}
\label{p0}
p_0(y)=\gamma{\gamma-\omega\over\omega}[e^{-(\gamma-\omega)y}
  -e^{-\gamma y}]\;.
\end{equation}
With this understanding, the proper normalization condition is
\begin{equation}
\int_0^y q(x,y)\,dx+c(y)^{-1}p_0(y)=1\;,
\end{equation}
which is indeed met!

As a last remark, notice that all of the above results are also valid when the
drift is {\em toward\/} the trap ($u<0$).  In this case the depletion zone
shrinks as the drift's strength increases, and it vanishes in the limit of
infinite drift.

\section{Shielding}
\label{sec:shielding} 

At first sight there seems to be a contradiction between the forward IPDF
(Eq.~\ref{ps}) and the concentration profile (Eq.~\ref{cs}).  How could it be
that the distribution of distances between nearest particles is independent of
location, while particles are sparser near the trap!?  ---The answer
is that the forward IPDF is a {\em conditional\/} probability---dependent on
the presence of the first particle at $x$.  Thus, the unexpectedly small
distance between particles in a sparse region is compensated by
the rare likelihood of finding such pairs of nearest particles in the first
place!  Nevertheless, one can't help wondering what kind of distribution would
explain the results of the previous section.  The answer turns out to be
surprisingly simple.

Suppose that the particles to the right of the trap are distributed just as in
equilibrium, with the exception of the nearest particle to the trap.  Assume
further that the nearest particle to the trap is at distance $z$, with
probability $p_0(z)$ (the same probability density as in Eq.~\ref{p0}).  In
this view, the effect of the trap is merely to create an empty gap between
itself and the first particle---a gap which increases as the drift $u$
approaches the critical value $v$.  [It follows from Eq.~(\ref{p0})
that the average distance to the first particle is $(2v-u)/(v-u)$.]  Another way
of looking at the suggested distribution is as if the nearest particle to the
trap {\em shields\/} all the other particles from the effects of
the trap!  The proposed distribution is illustrated in Fig.~3.

We now show that this model reproduces the results of Section~\ref{sec:trap}.
All we need to do is to derive the empty interval probability $E(x,y)$ implied
by the model, since everything else follows from it.  $E(x,y)$ can attain one
of three values---depending on the location of the endpoints of the
$(x,y)$ interval relative to the location of the nearest particle to the trap,
$z$:
\begin{equation}
E(x,y)=\left\{ \begin{array}{ll}
1 & \mbox{$x<y<z$}\;,\\
0 & \mbox{$x<z<y$}\;,\\
e^{-\gamma(y-x)} & \mbox{$z<x<y$}\;. \end{array}
\right.
\end{equation}
Therefore,
\begin{equation}
E(x,y)=\int_y^{\infty}p_0(z)\,dz+e^{-\gamma(y-x)}\int_0^xp_0(z)\,dz\;.
\end{equation}
The result is exactly as $E_s$ of Eq.~(\ref{Es.drift})!

\begin{figure} 
\centerline{\epsfxsize=5cm \epsfbox{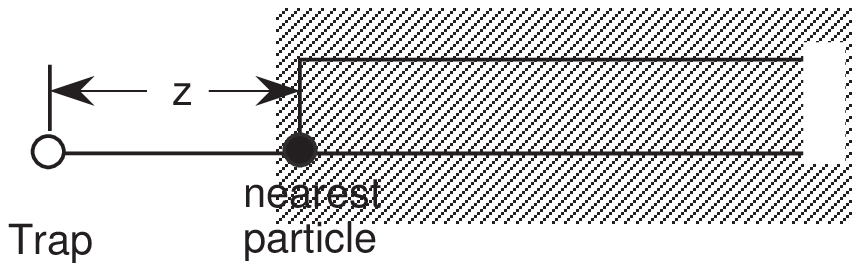}}
\noindent
{\small {\bf Figure~3}. Schematic illustration of the shielding effect:  The
particles in the shaded area are distributed randomly and independently from
each other, as in equilibrium.  The gap $z$ between the particles and the trap
follows the probability density distribution $p_0(z)$.}
\end{figure} 

So far, we have merely demonstrated that ``shielding" is {\em consistent\/}
with the exact results.  A complete proof requires showing that the full
hierarchy of
$n$-point density-density correlation functions are the same as postulated by
the  shielding model.  The problem of $n$-point correlation
functions can be formulated exactly within the framework of the IPDF method
\cite{Doering}.  With this method, it is rather straightforward to prove the
shielding effect~\cite{Fisher}.

\section{Reaction-diffusion equation}
\label{sec:mean-field}

Reaction-diffusion equations are one of the most valuable and widely used
approaches to kinetics of reaction processes.  The method posits
the existence of a {\em mesoscopic\/} length scale within which the system is
homogeneous and well mixed, and where the reaction rates can be accounted for
as in classical rate equations.  At longer length scales, variations in the
concentration, $c(x,t)$, give rise to diffusion, which is modeled by a
simple diffusion term.  For example, the reaction-diffusion equation that
applies to our model, in the absence of drift, is
\begin{equation}
\label{reac-diff}
{\partial c(x,t)\over\partial t}=D'{\partial^2c\over\partial x^2}+k_1c-k_2c^2\;,
\end{equation}
together with the boundary condition:
\begin{equation}
\label{c=0}
c(0,t)=0\;, 
\end{equation}
which is imposed by the trap at the origin.  $D'$ is the effective diffusion
constant, and $k_1$ and $k_2$ represent the effective rates of birth and
coalescence of particles, respectively.  

In spite of the popularity of reaction-diffusion equations
it is not often appreciated that they are mere idealizations, yielding only
approximate results.  The reason for this misconception is twofold: (1) few
exact solutions of reaction-diffusion models are available for comparison, and
(2) the reaction-diffusion equations themselves are usually complicated enough
that they cannot be solved in closed form, and so their consequences are not
always clear.  Diffusion-limited coalescence with a trap is a rare example
where the stationary state is known exactly, and where also the corresponding
reaction-diffusion equation may be solved exactly, at least in the case of zero
drift \cite{preprint}.  We wish to compare the two solutions.
 
The exact steady state for zero drift, needed for the comparison, may be
obtained from the results of Section~\ref{sec:trap}.  Taking the limit $w\to 0$
in Eqs.~(\ref{Es.drift}) and (\ref{cs}), we find
\begin{equation}
\label{Es}
E_s(x,y)=e^{-\gamma(y-x)}+\gamma(y-x)e^{-\gamma y}\;, 
\end{equation}
and
\begin{equation}
\label{cs-true}
c_s(x)=\gamma(1-e^{-\gamma x})\;.
\end{equation} 

Regarding the reaction-diffusion equation, Eq.~(\ref{reac-diff}), our first
concern is to identify the constants $D'$, $k_1$, and $k_2$.  We note first
that
\begin{equation}
D'=D\;,
\end{equation} 
since in
the absence of reactions (when $k_1,k_2=0$) the particles perform simple
diffusion, characterized by the diffusion constant $D$---the same $D$ as in
the hopping rate $D/a^2$ of the microscopic rules.  Furthermore, in the
continuum limit the model's only physical parameters relevant to its kinetics
are the diffusion constant
$D$ (of dimension $L^2/T$; $L\equiv{\rm length}$, and $T\equiv{\rm time}$) and
the birth rate $v$ (dimension
$L/T$). Therefore, the only way to produce the required dimensions of $k_1$
($1/T$) and
$k_2$ ($L/T$) is by having $k_1\sim v^2/D$ and $k_2\sim v$.
Finally, consider the stationary solution of Eq.~(\ref{reac-diff}) in an
infinite system without the trap: $c_{eq}=k_1/k_2$.  To conform with the
exact solution of~(\ref{ceq}) we must have $k_1/k_2=v/D$. Thus,
\begin{equation}
\label{k1/k2}
k_1=\beta{v^2\over D}\;;\qquad k_2=\beta v\;,
\end{equation}
where $\beta$ is a constant.

The stationary solution to
Eq.~(\ref{reac-diff}) with the trap---the boundary condition~(\ref{c=0})---is
\begin{equation}
\label{tanh}
{c_s(x)\over c_{\infty}}={3\over2}\tanh^2\Big(\sqrt{{k_1\over3D}}x
     +  \tanh^{-1}\sqrt{{1\over3}}\,\Big)-{1\over2}\;,
\end{equation}
where $c_{\infty}=k_1/k_2=v/D$ is the concentration of particles infinitely far
away from the trap. 
The concentration profile described by Eq.~(\ref{tanh}) looks
similar to the exact result of Eq.~(\ref{cs-true}).  One could now use
different criteria to determine the value of the fitting parameter $\beta$.
Demanding the correct asymptotic behavior far away from the trap;
$\lim_{x\to\infty}\ln[1-c_s(x)/c_{eq}]/x=-v/D$, we get $\beta=3/4$.
On the other hand, if we require the correct behavior close to the trap;
$(\partial c_s/\partial x)_{x=0}=(v/D)^2$, we get $\beta=9/4$.
Clearly, it is impossible to find a value of $\beta$ that reproduces the short
range behavior and the long range behavior simultaneously.  In Fig.~4 we show
the results of a least square fit in the range $0\leq (v/D)x\leq 5$, which is
achieved with $\beta=1.27\,$.
 
\begin{figure}
\centerline{\epsfxsize=6.5cm \epsfbox{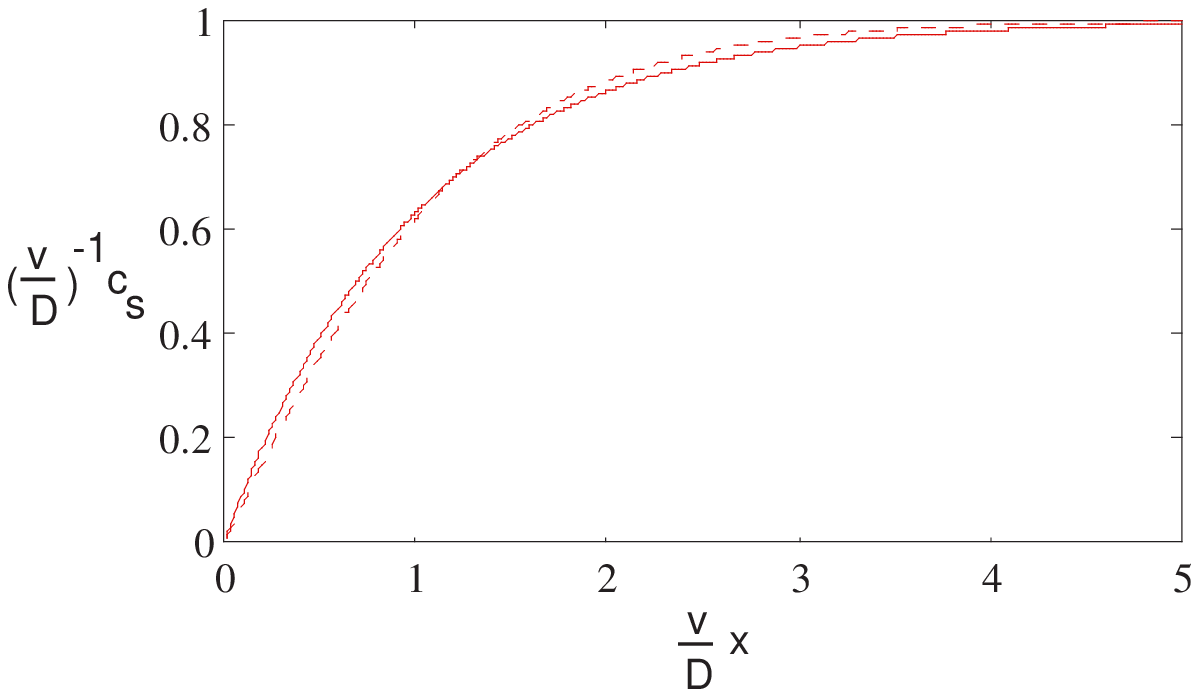}}
\noindent
{\small {\bf Figure~4}. Comparison of the exact analytic concentration profile
(solid line) to the prediction of the reaction-diffusion equation (broken
line).  Shown is the result of a best fit, achieved with $\beta=1.27\,$.}
\end{figure} 
 
\section{Discussion}
\label{sec:discussion}

We have solved the one-dimensional problem of diffusion-limited reversible
coalescence with a trap, and with a background drift, exactly.  The system 
exhibits a dynamical phase transition in its long-time asymptotic steady
state, which is controlled by the strength of the drift away from the trap.  As
the drift increases the depletion zone near the trap grows larger, until when
the drift crosses a critical threshold and the depletion zone becomes
infinitely wide; the system is then empty. 

A very intriguing prediction of the exact solution is the distribution of
distances between any given particle and its nearest neighbor to the right (the
forward IPDF), which turns out to be exactly the same as in equilibrium---as
in the system without the trap---and independent of the drift!  This is
explained by a ``shielding" effect:  the
nearest particle to the trap effectively shields the rest of the particles
from the trap's influence.  The trap and background drift affect only the width
of the gap between the trap and the nearest particle.

We have also contrasted the exact solution, in the absence of drift, with the
alternative, traditional approach of reaction-diffusion equations.  Our model
is a non-trivial example where both the true kinetics, as well as the 
solution to the model's reaction-diffusion equation, are known exactly.
Comparing the two solutions we were able to relate the
effective rates of the reaction-diffusion equation to the
microscopic rates of the underlying process, without appealing to
renormalization group techniques.
 
There remain several interesting open questions.  The exact solution of
Section~\ref{sec:trap} is basically an inspired guess.  It would be useful to
develop a more formal derivation method---one that would allow further
analysis of different generalizations of the problem at hand.  
Other open problems include the question of transient behavior.  For example,
beyond the transition point, when the drift exceeds the critical value of
$u_c=v$, it takes an infinite time for the system to achieve the empty steady
state.  What is the exact time dependence at different points of the phase
diagram?  ---What is the mean-field critical dimension beyond which the
reaction-diffusion equation becomes exact?  Previous work on a closely related
problem (Fisher waves~\cite{waves}) has suggested that $d_c=3$, but it was based
on computer simulations of an ``infinite" system and the results were
controversial.  The bounding of our present model by the trap at the origin may
offer a better alternative for numerical studies.  ---What is the prediction of
the reaction-diffusion equation formulation when drift is included?  How well is
the dynamical phase transition captured in this approximation?  These and
similar issues will be the subject of future research.

\acknowledgments
I thank L.~Glasser, P.~Krapivsky, V.~Privman, S.~Redner, and L.~Schulman for
valuable discussions.


\end{multicols}


\begin{thebibliography}{99}
\bibitem{remark} Our notation here differs from previous work: we take the
birth rate to be $v/a$ rather than $v/2a$, to achieve a more aesthetic form of
the final result.
\end{thebibliography}
\end{document}